\documentclass[twocolumn,prx, superscriptaddress, longbibliography, floatfix,aps]{revtex4-1}
\usepackage[T1]{fontenc}
\usepackage{graphicx}
\usepackage{amsfonts}
\usepackage{amsmath}
\usepackage{amssymb}
\usepackage[dvipsnames]{xcolor}
\usepackage[colorlinks,linkcolor=Blue,citecolor=Blue]{hyperref}
\usepackage{braket}
\usepackage{microtype}
\usepackage{bbold}

\begin{document}

\title{Berry phase effects on the transverse conductivity of Fermi surfaces and their detection via spin qubit noise magnetometry}

\author{Mark Morgenthaler}
\affiliation{Institut f{\"u}r Theoretische Physik, Universit{\"a}t Leipzig, Br{\"u}derstra{\ss}e 16, 04103, Leipzig, Germany}

\author{Inti Sodemann Villadiego}
% \email[]{sodemann@uni-leipzig.de}
\affiliation{Institut f{\"u}r Theoretische Physik, Universit{\"a}t Leipzig, Br{\"u}derstra{\ss}e 16, 04103, Leipzig, Germany}

\date{\today}

\begin{abstract}
The quasi-static transverse conductivity of clean Fermi liquids at long wavelengths displays a remarkably universal behaviour: it is determined solely by the radius of curvature of the Fermi surface and does not depend on details such as the quasi-particle mass or their interactions. Here we demonstrate that Berry phases do not alter such universality by directly computing the transverse conductivity of two-dimensional electronic systems with Dirac dispersions, such as those appearing in graphene and its chiral multilayer variants. Interestingly, however, such universality ceases to hold at wave-vectors comparable to the Fermi radius, where Dirac fermions display a vividly distict transverse conductivity relative to parabolic Fermions, with a rich wave-vector dependence that includes divergences, oscillations and zeroes. We discuss how this can be probed by measuring the $T_1$ relaxation time of spin qubits, such as NV centers or nuclear spins, placed near such 2D systems.
\end{abstract}

\maketitle

\textit{\bfseries \color{blue} Introduction.} Demonstrating the existence of observables in metals that are highly insensitive to details (i.e ``universal'') is of great interest because they can serve as the basis to devise robust experimental probes to characterise them. A paradigmatic example of such universal observable is the period of quantum oscillations in response to magnetic fields, which is routinely employed to measure the area of the Fermi surface \cite{Abrikosov_2017}. Another less well-known universal observable is the low temperature real part of the transverse conductivity in the quasi-static ($\omega \ll v_F q$) and clean long-wavelength limit ($l_{\rm mfp} \ll  q \ll k_F$), which is entirely determined by the geometric radius of curvature of the Fermi surface \cite{Pines_2018, Khoo_2021} and for a two-dimensional Fermi surface approaches the value:
\begin{equation}\label{eq:conductivity_intro}
    \mathrm{Re} \,  \sigma_\perp(q, \omega)\to \frac{e^2}{h} \frac{k_F}{q} \ .
\end{equation}
The above is universal in the sense that it only depends on the radius of the Fermi surface (or its local radius of curvature when it is not a circle) and not on the kinetic details of the energy dispersion or the electron-electron interactions, i.e. it is the same for the Fermi gas and the Fermi liquid \cite{Khoo_2021, Morgenthaler_2024}. Moreover, while difficult to probe in traditional transport experiments, the above regime can be accessed by measuring the relaxation times of spin qubits such as NV centers or nuclear spins \cite{Kolkowitz_2015, Agarwal_2017, Ariyaratne_2018, Khoo_2021, Khoo_2022}.

In this study we extend the scope of the above universality by demonstrating that the the transverse conductivity for Fermi surfaces with non-trivial Berry phases approaches the same as in Eq.\eqref{eq:conductivity_intro}. We will do this by computing the quasi-static transverse conductivities and their contributions to the $T_1$ time in 2D Dirac fermions of graphene and their $N$-layer chiral variants (see e.g. \cite{Min_2008}). However, we will also show that when the wave-vector $q$ becomes comparable to $k_F$, the chiral fermions have drastically different transverse conductivities compared to electrons without Berry phases. For example, the transverse conductivity for the standard Dirac fermion of monolayer graphene diverges  when $q \to 2k_F$, in contrast to parabolic fermions where the conductivity vanishes for $q \to 2k_F$. This divergence occurs for any odd-layer chiral fermions (i.e. rhombohedral tri-layer or penta-layer graphene) but not for even $N$ (i.e. Bernal bilayer graphene). We have also found that in these multi-layer chiral fermions the conductivity oscillates and displays approximately $N/2$ zeroes as a function of $q$. Remarkably, despite all these differences in their detailed wave-vector dependence, the integral of the conductivity which determines the $1/T_1$ decay rate of a spin qubit at distances smaller than the Fermi wave-length, has the exact same value for a trivial parabolic fermion or any chiral $N$-layer fermion, except for $N=1$ (i.e. mono-layer graphene), for which it is twice as large as a trivial parabolic fermion.

\medskip

\textit{\bfseries \color{blue} General Setting.} 
We consider a transverse oscillating electric field $\mathbf{E}(\mathbf{r}, t) = E_{q, \omega} \mathrm{e}^{\mathrm{i} (qx - \omega t)} \mathbf{\hat{y}} + \mathrm{c.c.}$ pointing in $y$-direction with wave vector $\mathbf{q} = q \mathbf{\hat{x}}$ pointing in $x$-direction which couples to the electron Hamiltonian via a vector potential $\mathbf{E}(\mathbf{r}, t) = - \partial \mathbf{A}(\mathbf{r}, t)/\partial t$ as follows ($e = \hbar = 1$):
\begin{align}
    H = h\bigl( \mathbf{p} - \mathbf{A}(\mathbf{r}, t) \bigr) \ ,
\label{eq:general_hamiltonian}
\end{align}
where $h$ is a matrix which we will specify for different systems of interest in the coming sections. The above captures the orbital coupling electrons to the fields, the spin coupling will be discussed in a later section. The real part of the transverse conductivity  obtained from the current-current response function, is given by (see Appendix \ref{appendixA}):
\begin{align}
\label{eq:general_transverse_conductivity}
    \mathrm{Re} \, \sigma_\perp(q, \omega) &= \frac{\pi}{4 \omega} \int \frac{\mathrm{d}^2 k}{(2\pi)^2} \sum_{s_1, s_2} \frac{n_{\mathbf{k} - \mathbf{q}, s_1} - n_{\mathbf{k}, s_2}}{\omega + \varepsilon_{\mathbf{k} - \mathbf{q}, s_1} - \varepsilon_{\mathbf{k}, s_2} + \mathrm{i} \eta} \\ 
    & \times \left\vert \bra{\mathbf{k} - \mathbf{q},s_1} \bigg( v_y(\mathbf{k} - \mathbf{q}) + v_y(\mathbf{k}) \bigg) \ket{\mathbf{k},s_2} \right\vert^2 \notag,
\end{align}
where $\ket{\mathbf{k},s}$ is the band eigen-ket from $h(\mathbf{k})$ with eigen-energy $\varepsilon_{\mathbf{k}, s}$ and Fermi-Dirac occupation $n_{\mathbf{k}, s}$, $v_y(\mathbf{k}) = \partial h(\mathbf{k})/\partial k_y$ is the velocity operator along $y$.

%, and we take the limit $\eta \to 0^+$.

The above conductivity is related to the current fluctuations by the fluctuation-dissipation theorem \cite{Giuliani_Vignale_2005}, and such fluctuations act as a source of magnetic field fluctuations, leading to a relation between the above conductivity and the fluctuations of magnetic field given by \cite{Casola_2018, Langsjoen_2012, Khoo_2021}:
\begin{align}\label{eq:general_magnetic_noise}
    \chi_{B_z, B_z}(z, \omega) = \frac{\mu_0 \omega^2}{4} \int \frac{\mathrm{d}^2 q}{(2\pi)^2} \mathrm{e}^{-2 q z} \mathrm{Re} \, \sigma_\perp(q, \omega) \ ,
\end{align}
where $\mu_0$ is the vacuum magnetic permeability, and $z$ denotes the height above the 2D system where the noise is locally measured. This noise is proportional to the $1/T_1$ decay rate of a spin magnetic moment located at a distance $z$ above the two-dimensional system of interest (i.e. in the case of a nuclear spin embedded in the 2D system itself, one can take $z \approx 0$). In the following we will compute the quasi-static transverse conductivity and magnetic noise of different non-interacting fermionic systems based on these formulae, focusing on their quasi-static limits (i.e. $\omega \to 0$ while $q$ is finite), which are relevant for the nuclear and NV-center $1/T_1$ decay rates \cite{Casola_2018, Langsjoen_2012, Khoo_2021}.

\medskip

% \noindent \textit{\bfseries \color{blue} Orbital Contributions.}
\textit{\bfseries \color{blue} Galilean \& Dirac Fermions.}
We begin by considering an isotropic electronic dispersion without Berry phases, such as that arising in a tight-binding model without spin-orbit coupling and a single orbital per unit cell. Namely in Eq.\eqref{eq:general_hamiltonian}, we take 

\begin{align}
    h({\bf p})=\varepsilon(\vert \mathbf{p} \vert) \ ,
\label{eq:Galilean_hamiltonian}
\end{align}

\noindent where $\varepsilon$ is any function of the magnitude of momentum, including the case of a parabolic dispersion, $\varepsilon(\vert \mathbf{p} \vert)=\mathbf{p}^2/2m$, and thus we will refer to this case as ``Galilean fermions'', but the formulae below apply for any $\varepsilon(\vert \mathbf{p} \vert)$. The system has a circular Fermi surface of radius $k_F$, defined by $\varepsilon(\vert \mathbf{p} \vert)=\varepsilon(k_F)$. From Eq.\eqref{eq:general_transverse_conductivity}, the quasi-static transverse conductivity can be shown to be: 

%conductivity for an isotropic dispersion $\varepsilon(\mathbf{p}) = \varepsilon(\vert \mathbf{p} \vert)$ in the quasi-static limit can be calculated to be

%we consider a Hamiltonian
%\begin{align}
%    H = \varepsilon\bigl( \mathbf{p} - \mathbf{A}(\mathbf{r}, t) \bigr) \quad \left(\mathrm{e.g.} \ H = \frac{\bigl( \mathbf{p} - \mathbf{A}(\mathbf{r}, t) \bigr)^2}{2m} \right) %\ .
%\label{eq:Galilean_hamiltonian}
%\end{align}

\begin{align}
    \sigma_\perp^G(q)\equiv \lim_{\omega \to 0} \mathrm{Re}\, \sigma_\perp^G(q, \omega) = \Theta(2k_F - q) \frac{e^2}{h} \frac{k_F}{q} \sqrt{1 - \frac{q^2}{4k_F^2}} \ ,
\label{eq:Galilean_conductivity}
\end{align}
where the superscript $G$ stands for "Galilean fermions". The corresponding quasi-static magnetic noise from Eq.\eqref{eq:general_magnetic_noise} for $z \to 0$, is:

\begin{align}\label{eq:Galilean_magnetic_noise}
    \chi^G \equiv \lim_{\omega \to 0}\chi^G_{B_z, B_z}(0,\omega)= \frac{e^2}{h} \frac{\mu_0 \omega^2 k_F^2}{32\pi} .
\end{align}

\noindent Remarkably, we see that the quasi-static transverse conductivity and the corresponding magnetic noise only depend on the size of the Fermi surface and not on any details of the electronic dispersion $\varepsilon(\vert \mathbf{p} \vert)$.

We now consider the case of a massive 2D Dirac fermion, where in Eq.\eqref{eq:general_hamiltonian} we would have:
\begin{align}
    h(\mathbf{p})= v (p_x  \tau_x+p_y  \tau_y) + \Delta \tau_z
\label{eq:Dirac_hamiltonian}
\end{align}
where $v$ is the Dirac velocity, $\Delta$ the gap, and $\tau_{x,y,z}$ are Pauli matrices. From Eq.\eqref{eq:general_transverse_conductivity} we find that:
\begin{align}
    \sigma_\perp^D(q)=\lim_{\omega \to 0} \mathrm{Re}\, \sigma_\perp^D(q, \omega) = \sigma_\perp^G(q) + \delta \sigma_\perp^D(q) \ ,
\label{eq:Dirac_conductivity}
\end{align}
where
\begin{align}\label{eq:magnetic_moment_correction}
    \delta \sigma_\perp^D(q) = \frac{\Theta(2k_F - q)}{4} \frac{e^2}{h} \frac{q}{k_F} \frac{1}{\sqrt{1 - \frac{q^2}{4k_F^2}}} \ .
\end{align}
And the corresponding magnetic noise is:
\begin{align}\label{noiseDirac}
    \chi^D\equiv\lim_{\omega \to 0}\chi^D_{B_z, B_z}(0,\omega) = \frac{e^2}{h} \frac{\mu_0 \omega^2 k_F^2}{16 \pi} = 2 \chi^G \ .
\end{align}

\noindent Remarkably, again, the above results only depend on the Fermi radius and are independent of the gap $\Delta$.

%effective Compton wave-vector $q_\Delta\equiv \Delta/v$. 

\begin{figure}[t]
    \includegraphics[width=0.48\textwidth]{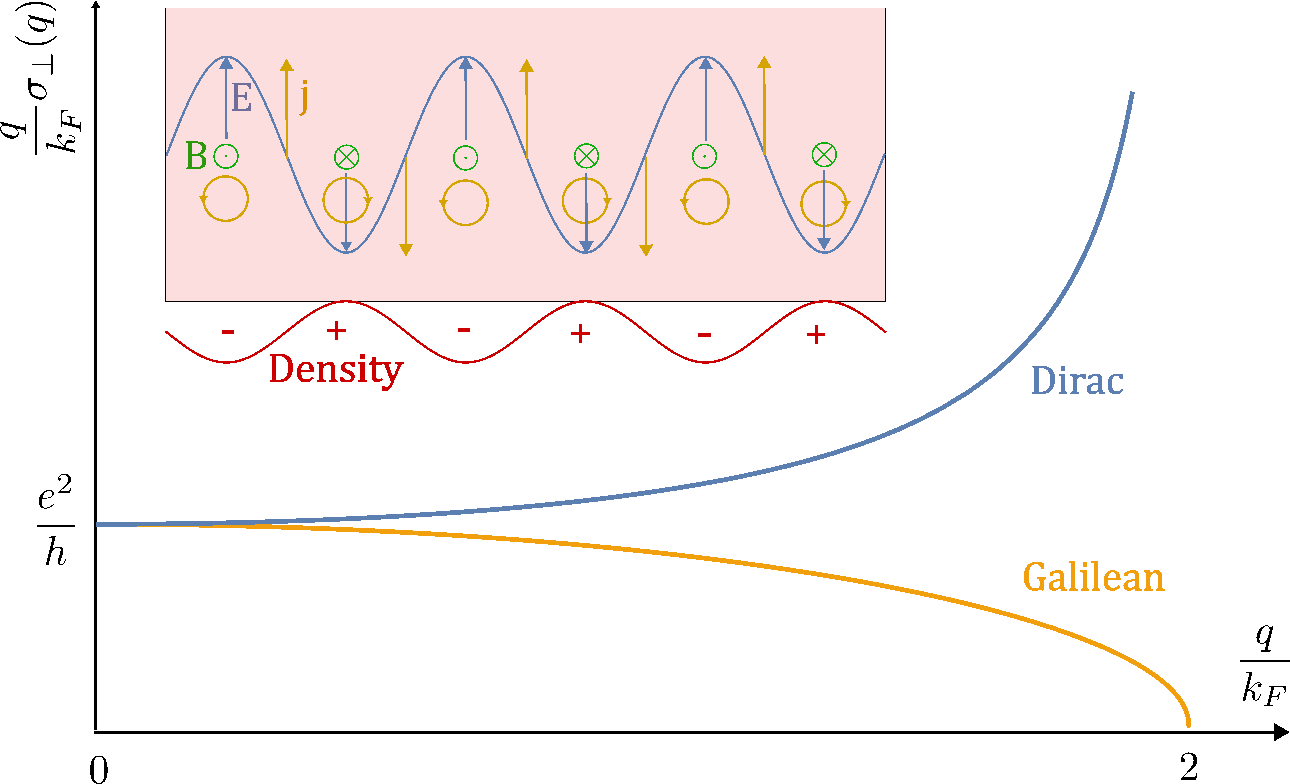}
    \caption{Quasi-static transverse conductivities of Galilean and Dirac  fermions as a function of the wavevector $q$, which are identical for $q \to 0$ but differ for $q \to 2k_F$. \textbf{Inset:} Depiction of the additional current induced by the orbital magnetic moment in Dirac fermions. Transverse electric fields (blue arrows) are accompanied by oscillating out-of-plane magnetic fields (green symbols), which induce a density oscillation (bottom red ``$\pm$'' symbols) because the Dirac particles carry an out-of-plane magnetic moment (orange circulating symbols), leading to an additional transverse current (orange arrows) peaked at the nodes where the magnetic field changes sign.}
    \label{fig:Dirac_conductivity}
\end{figure}

At first glance the difference between the Galilean and Dirac fermions appears at odds with the principle that their behavior should be the same in the non-relativistic limit: $q,k_F \ll \Delta/v$ (i.e. Bohr's correspondence principle). This discrepancy can be reconciled by noticing that the correct low energy approximation to the Dirac Hamiltonian includes also an orbital magnetic moment:
\begin{align}\label{eq:orbital_moment_hamiltonian}
    H = \frac{\bigl( \mathbf{p} - \mathbf{A}(\mathbf{r}, t) \bigr)^2}{2m} -\frac{(\nabla \times \mathbf{A}(\mathbf{r}, t))_z}{2m} \ ,
\end{align}
where $m=\Delta/v^2$. This additional orbital moment coupling, which can also be viewed as a Berry phase effect \cite{Xiao_2010}, gives rise to a correction to the current operator of Galilean fermions. Remarkably, as we show in Appendix \ref{appendixB}, the above parabolic Hamiltonian with the orbital moment term has exactly the same quasi-static transverse conductivity and same magnetic field noise as the Dirac fermion without any approximations.  This also clarifies the origin of enhanced current fluctuations of Dirac fermions: they arise from the additional density fluctuations induced by the magnetic field coupling to the orbital moment, giving rise to a larger transverse conductivity as depcícted in Fig.\ref{fig:Dirac_conductivity}. 

The above should not be viewed as a statement that the Hamiltonian of Eq.\eqref{eq:Galilean_hamiltonian} is incorrect or missing the extra term in Eq.\eqref{eq:orbital_moment_hamiltonian}. The two cases correspond to two different physical settings. Namely, Eq.\eqref{eq:Galilean_hamiltonian} arises from a trivial band with a single orbital per unit cell and no Berry phases, while the Dirac fermion or Eq.\eqref{eq:orbital_moment_hamiltonian} should be viewed as arising from a model with at least two orbitals per unit cell and non-trivial Berry phases. For this reason, we will refer to Eq.\eqref{eq:orbital_moment_hamiltonian} as the ``low-energy Dirac fermion'', to distinguish it from the case of the strictly parabolic Galilean fermion from Eq.\eqref{eq:Galilean_hamiltonian}.

%The details for the calculation of the conductivity can be found in Appendix \ref{appendixB} and we find that the conductivity for the low-energy Dirac fermion is exactly identical to the one for the Dirac fermion (with no approximations needed), with the additional correction to the Galilean conductivity arising entirely due to the orbital magnetic moment.

% \begin{figure}[t]
% \includegraphics[width=0.48\textwidth]{PRL_Pictures/Galilean_Dirac_conductivity.png}
% \caption{Quasi-static transverse conductivities of Galilean and Dirac  as a function of the wavevector $q$. One can see that the behaviour in the limit of long wavelengths ($q \to 0$) is the same, but changes as $q \to 2k_F$.}
% \label{fig:Dirac_conductivity}
% \end{figure}

% \begin{figure}[t]
% \includegraphics[width=0.48\textwidth]{PRL_Pictures/magnetic_moment_current.png}
% \caption{Schematic of the additional current induced by the orbital magnetic moment. Transverse electric fields (blue arrows) are accompanied by oscillating out-of-plane magnetic fields (green symbols), which induce a density oscillation (bottom red ``$\pm$'' symbols) since the Dirac particles carry an out-of-plane magnetic moment (orange circulating symbols), leading to an additional transverse current (orange arrows) peaked at the nodes where the magnetic field changes sign.}
% \label{fig:OrbitalMagneticMoment_current}
% \end{figure}

\medskip

\textit{\bfseries \color{blue} Multilayer Graphene Chiral Fermions.} Now consider the low energy Hamiltonians arising $N$-layer chiral stacking of graphene, which can be approximated as follows \cite{Min_2008}:

\begin{align}
    h({\bf p}) = 
\begin{pmatrix}
 \Delta  & \alpha_N (p_x-i p_y)^N \\
 \alpha_N (p_x+i p_y)^N & -\Delta  \\
\end{pmatrix} ,
\label{eq:multilayer_hamiltonian}
\end{align}

\begin{figure}[t]
    \includegraphics[width=0.48\textwidth]{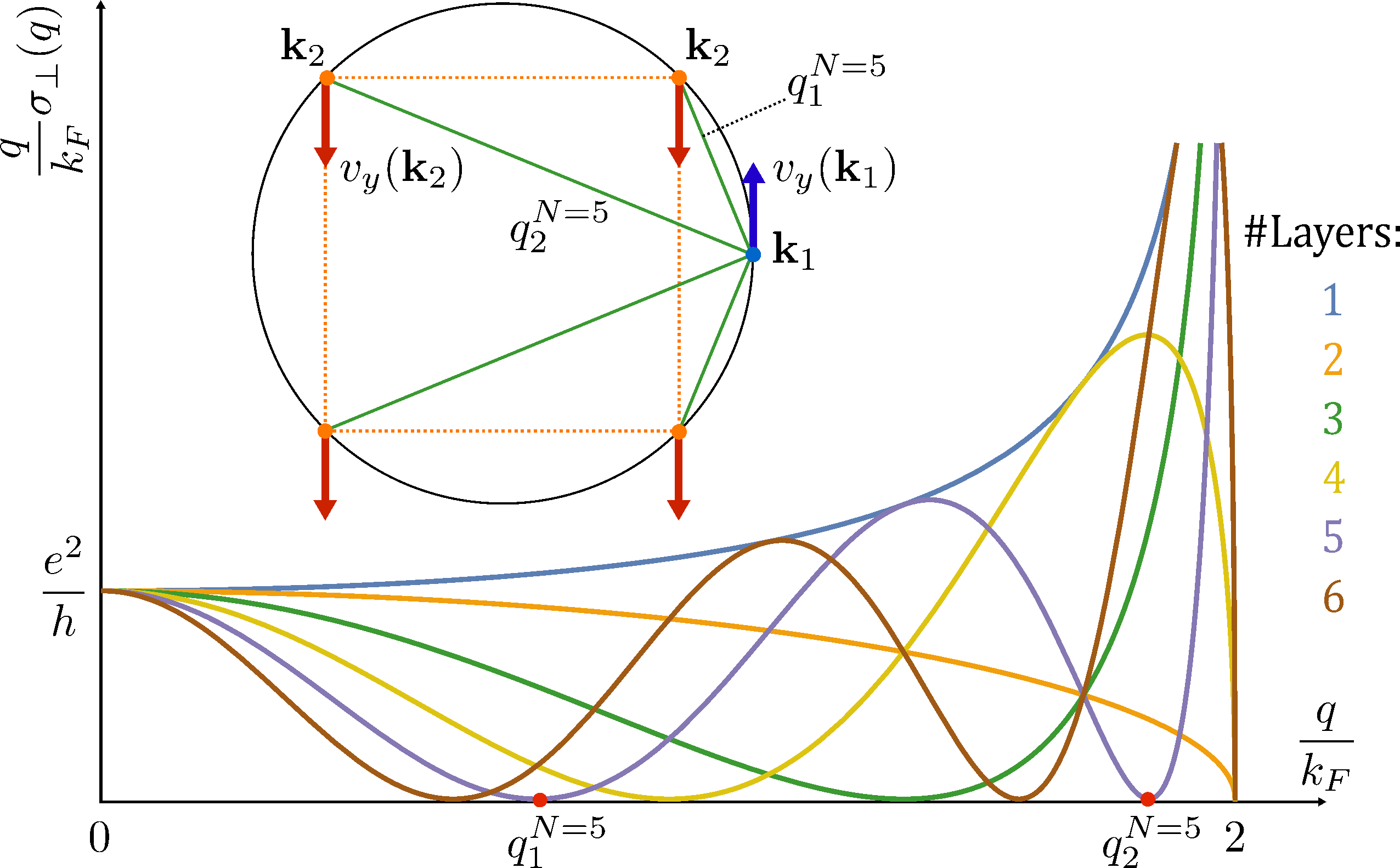}
    \caption{Quasi-static transverse conductivity for $N$-layer graphene chiral fermions. For $N>1$, the conductivity oscillates between zero and the value for usual Dirac fermions ($N=1$ in blue). The conductivity diverges for odd $N$ and vanishes for even $N$ as $q \to 2k_F$. Interestingly, the bilayer conductivity ($N=2$) is identical to the Galilean conductivity. \textbf{Inset:} Depiction of the geometry of the zeros of the conductivity for $N = 5$ (pentalayer graphene). The circle is the Fermi surface. The state ($\mathbf{k}_1$ with blue dot), scatters by wave-vector $\mathbf{q}$ to another state (orange dot with $\mathbf{k}_2 = \mathbf{k}_1 + \mathbf{q}$) on the Fermi surface. The zeros arise when the velocity matrix elements from Eq.\eqref{eq:velocity_operator_zeros} vanishes for such transition. These zeroes can be found by fitting an $N-1$ sided regular polygon inside the Fermi surface, i.e. a square for $N = 5$ (orange dashed line).}
    \label{fig:multilayer_conductivity}
\end{figure}

%\begin{align}
  %  H = \alpha_N \begin{pmatrix} \mathrm{Re}\, \xi_N\bigl( \mathbf{p} - \mathbf{A}(\mathbf{r}, t) \bigr) \\ \mathrm{Im}\, \xi_N\bigl( \mathbf{p} - \mathbf{A}(\mathbf{r}, t) \bigr) \end{pmatrix} \cdot \boldsymbol{\tau} + \Delta \tau_z \ ,
%\label{eq:multilayer_hamiltonian}
%\end{align}
%with
%\begin{align}\label{eq:pseudo_spin}
%    \xi_N(\mathbf{p}) = (p_x + \mathrm{i} p_y)^{N} \ , \quad \boldsymbol{\tau} = \begin{pmatrix}
%        \tau_x \\ \tau_y
%    \end{pmatrix} \ .
%\end{align}
% \begin{align}
%     H = \boldsymbol{\xi}\bigl( \mathbf{p} - \mathbf{A}(\mathbf{r}, t) \bigr) \cdot \boldsymbol{\sigma} + \Delta \sigma_z \ ,
% \label{eq:multilayer_hamiltonian}
% \end{align}
% with
% \begin{align}\label{eq:pseudo_spin}
%     \boldsymbol{\xi}(\mathbf{p}) = \alpha_N \begin{pmatrix} \mathrm{Re}\, (p_x + \mathrm{i} p_y)^N \\ \mathrm{Im}\, (p_x + \mathrm{i} p_y)^N \end{pmatrix} \ .
% \end{align}
\noindent Where $\Delta$ and $\alpha_N$ are constants. The case $N = 1$ corresponds to the already considered Dirac fermion, while for $N > 1$, we obtained the following analytic expression for their quasi-static conductivity (see Appendix \ref{appendixC}):
\begin{align}\label{eq:multilayer_conductivity}
    \sigma_\perp^N(q) \equiv \lim_{\omega \to 0} \mathrm{Re}\, \sigma_\perp^N(q, \omega) = \sigma_\perp^D(q) P_N(q) \ ,
\end{align}
where
\begin{align*}
    P_N(q) = \frac{(-1)^{N-1}}{4} &\prod_{n=0}^{N-2} \left( \frac{q^2}{k_F^2} - 4 \sin^2\left( \frac{(2n + 1) \pi}{2 (N-1)} \right) \right) \ .
\end{align*}
This conductivity, plotted in Fig. \ref{fig:multilayer_conductivity}, exhibits several remarkable properties. First it is independent of the gap $\Delta$ and only depends $k_F$ and the index $N$. We also see in Fig. \ref{fig:multilayer_conductivity}, that this conductivity never exceeds that of the standard Dirac fermion ($N=1$) for any $N$. Additionally, we see that the transverse conductivity oscillates and has $N/2$ and $(N-1)/2$ zeros at intermediate wave vectors for even and odd values of $N$, respectively. Additionally, for odd $N$ the conductivity diverges at the threshold $q \to 2k_F$, while for $N$ even it vanishes. Even more surprisingly, despite the completely distinct wave-vector dependences for different $N$, the integral over wave-vectors which determines the magnetic noise at the location of the 2D system (i.e. $z \to 0$ in Eq.\eqref{eq:general_magnetic_noise}), is exactly the same for all $N>1$ and identical to that of a Galilean fermion, but it is twice such value for the standard Dirac fermion, namely:

% As can be seen in figure \ref{fig:multilayer_conductivity}, the additional polynomial $P_N$ causes the conductivity to oscillate between zero and the Dirac conductivity. Surprisingly, these oscillations do not change the area under the graph (corresponding to the magnetic noise), resulting in
\begin{align}\label{eq:N_layer_magnetic_noise}
    \chi^N \equiv \lim_{\omega \to 0} \chi_{B_z, B_z}^N(0, \omega) = \begin{cases}
        2 \chi^G \ &, \ N = 1 \\
        \chi^G \ &, \ N \geq 2
    \end{cases}
\end{align}

% Note that since the Berry phase is known to be $\phi_\mathrm{Berry} = N \pi$, one can find a curious relation between the behaviour of the conductivity and the Berry phase:
% \begin{align}
%     \frac{\phi_\mathrm{Berry}}{\pi} = 2N_\mathrm{Zeros} + N_\mathrm{Divergences} \ ,
% \label{eq:berry_phase_relation}
% \end{align}
% where $N_\mathrm{Zeros}$ and $N_\mathrm{Divergences}$ denote the number of (unique) zeros and divergences of the conductivity respectively.

% We can visualise the zeros of the conductivity by looking at the velocity operator
% \begin{align}
%     v_y(\mathbf{k}) = N \alpha_N \begin{pmatrix} -\mathrm{Im}\, \xi_{N-1}(\mathbf{k}) \\ \mathrm{Re}\, \xi_{N-1}(\mathbf{k}) \end{pmatrix} \cdot \boldsymbol{\tau} \ ,
% \label{eq:NLayer_velocity_operator}
% \end{align}
% as the conductivity depends on its matrix elements (see equation \eqref{eq:general_transverse_conductivity}). One finds that {\color{red} (comment on + in ket)}

The origin of the zeroes of the transverse conductivity can be traced back to special wave-vectors for which there is a vanishing of the velocity matrix elements in Eq.\eqref{eq:general_hamiltonian} associated with the scattering between two states at momenta $\mathbf{k}_1$ and $\mathbf{k}_2$ (with $\mathbf{q} = \mathbf{k}_2 - \mathbf{k}_2$) located at the Fermi surface (see inset of Fig. \ref{fig:multilayer_conductivity}). They can be found by solving the following equation:
\begin{align}\label{eq:velocity_operator_zeros}
    \bra{\mathbf{k}_1, +} \bigg( v_y(\mathbf{k}_1) + v_y(\mathbf{k}_2) \bigg) \ket{\mathbf{k}_2, +} = 0 \ ,
\end{align}
which leads to the following condition (see inset of Fig. \ref{fig:multilayer_conductivity}) $({k}_{1x} + \mathrm{i} {k}_{1y})^{N-1} = -({k}_{2x} + \mathrm{i} {k}_{2y})^{N-1}$, whose solutions lead to the following expression for the conductivity zeroes:
\begin{align}\label{eq:multilayer_zeros}
    q_m = k_F \left\vert \mathrm{e}^{\mathrm{i} \frac{(2m + 1) \pi}{N-1}} - 1 \right\vert \ , \quad m = 0, \dots, N-1.
\end{align}

% \begin{align}\label{eq:zeros_condition}
%     \xi_{N-1}(\mathbf{k}_1) = -\xi_{N-1}(\mathbf{k}_2) \ .
% \end{align}

%Conversely, the conductivity will be ``maximal'' (i.e. correspond to the Dirac conductivity) exactly when
%\begin{align}\label{eq:maxima_condition}
%    ({k}_{1x} + \mathrm{i} {k}_{1y})^{N-1} = ({k}_{2x} + \mathrm{i} {k}_{2y})^{N-1} \ .
%\end{align}

% \begin{figure}[t]
% \includegraphics[width=0.48\textwidth]{PRL_Pictures/Multilayer_conductivity.png}
% \caption{Quasi-static transverse conductivity for the multilayer chiral fermion. For $N>1$, the conductivity oscillates between zero and the Dirac conductivity (blue). The conductivity diverges for odd $N$ and vanishes for even $N$ as $q \to 2k_F$. Interestingly, the bilayer conductivity ($N=2$) is identical to the Galilean conductivity.}
% \label{fig:multilayer_conductivity}
% \end{figure}

% \begin{figure}[t]
% \includegraphics[width=0.48\textwidth]{PRL_Pictures/conductivity_zeros.png}
% \caption{Visualising the zeros of the conductivity for the chiral fermion with $N = 5$ relevant for chiral pentalayer graphene. The circle corresponds to the Fermi surface. Fixing an arbitrary initial state $\mathbf{k}_1$ (blue dot) on the Fermi surface, the electric field with wave-vector $\mathbf{q}$ scatters this to another state $\mathbf{k}_2 = \mathbf{k}_1 + \mathbf{q} $ (orange dot) on the Fermi surface (as $\omega = 0$). The zeros of the conductivity arise when the wave-vector $\mathbf{q}$ is such that the velocity matrix elements from Eq.\eqref{eq:velocity_operator_zeros} vanish. These special points can be found by fitting an $N-1$ sided regular polygon inside the Fermi surface, i.e. a square for $N = 5$ (orange dashed line).}
% \label{fig:conductivity_zeros}
% \end{figure}

\medskip

\textit{\bfseries \color{blue} Spin Magnetic Moment Corrections.}\label{sec:vacuum_magnetic_moment} To account for contributions from the microscopic spin of the electrons we must also add a Zeeman coupling of the form:
\begin{align}\begin{aligned}
    \delta H = -\mu_\mathrm{s} (\nabla \times \mathbf{A}(\mathbf{r}, t))_z \sigma_z
\label{eq:vacuum_moment_hamiltonian_correction}
\end{aligned}\end{align}
where $\mu_\mathrm{s}= 1/(2m_0)$ is the spin magnetic moment and $\sigma_z$ is the spin Pauli matrix. The orbital contributions to the transverse conductivity described in the previous sections were computed for spinless electrons, and thus they acquire an additional trivial factor of $2$ due to spin degeneracy. But in addition to such trivial correction, there is also a correction to the current operator which leads to a non-trivial wave-vector dependent correction to the transverse conductivity. The details of the derivation are presented in Appendix \ref{appendixD}, which lead to  additional contributions which we will present next for each of the cases.

For the Galilean fermion, the Zeeman induced correction to the conductivity is given by

\begin{align}\label{eq:Galilean:spin_correction}
    \lim_{\omega \to 0} \delta \sigma_\perp^{G,{\rm spin}}(q, \omega) = 8 \delta \sigma_\perp^D(q) \frac{k_F^2}{v_F^2} \mu_s^2 = 2 \delta \sigma_\perp^D(q) \frac{m^2}{m_0^2} \ ,
\end{align}
%\begin{align}
%    \lim_{\omega \to 0} \delta \sigma_\perp^{G,{\rm spin}}(q, \omega) = 8M(q) \frac{k_F^2}{v_F^2} \mu_\mathrm{vacuum}^2 = 2M(q) \frac{m^2}{m_0^2} \ ,
%\end{align}
where the last equality holds only for strictly parabolic fermions, we have set $v_F = \partial_k \varepsilon_k \bigr\vert_{k = k_F}$, and used $\delta \sigma_\perp^D$ from Eq.\eqref{eq:magnetic_moment_correction}. The magnetic noise for the entire conductivity (including spin degeneracy) then becomes
\begin{align}
    \chi_\mathrm{total}^G = 2\chi^G \left( 1 + \frac{m^2}{m_0^2} \right) \ .
\end{align}
The same correction term will also arise for the low-energy Dirac fermion (i.e. the one described by Eq.\eqref{eq:orbital_moment_hamiltonian}), though the total magnetic noise will differ, due to the extra contribution of the orbital correction to the magnetic moment from Eq.\eqref{eq:orbital_moment_hamiltonian}, leading to a total noise of:
\begin{align}\label{eq:LED_total_magnetic_noise}
    \chi_\mathrm{total}^{LD} = 2\chi^G \left( 2 + \frac{m^2}{m_0^2} \right) \ .
\end{align}

\noindent For the standard Dirac fermion from Eq.\eqref{eq:Dirac_hamiltonian}, the correction to the transverse conductivity will be
\begin{align}
    \lim_{\omega \to 0} \delta \sigma_\perp^{D, \mathrm{spin}}(q, \omega) = 2 \delta \sigma_\perp^D(q) \frac{\Delta^2 + v^2 k_F^2 \left( 1 - \frac{q^2}{4k_F^2} \right)}{m_0^2 v^4} \ .
\end{align}
Notice that in contrast to the orbital contributions, the above expressions for the Zeeman contributions depend explicitly on the gap of the Dirac fermions, although, as expected, they reduce to the expression for the low-energy Dirac fermion from Eq.\eqref{eq:LED_total_magnetic_noise} in the non-relativistic $\Delta \gg v k_F$, $\Delta \gg v q$.
% which now only agrees with the low-energy Dirac fermion up to leading orders in $\Delta$ (as opposed to them agreeing exactly as before). 
The total magnetic noise will then be given by
\begin{align}
    \chi_\mathrm{total}^D = 2\chi^G \left( 2 + \frac{\Delta^2 + \frac{k_F^2 v^2}{4}}{m_0^2 v^4} \right) \ .
\end{align}
On the other hand, for the chiral fermions from Eq. \eqref{eq:multilayer_hamiltonian}, the spin correction to the transverse conductivity is:
\begin{align}\label{eq:multilayer_spin_correction}
    \lim_{\omega \to 0} \delta \sigma^{N, \mathrm{spin}}_\perp(q, \omega) = 2 \delta \sigma_\perp^D(q) \frac{1}{N^2} \frac{\Delta^2 + \alpha_N^2 k_F^{2N} (1-\beta_N)}{m_0^2 \alpha_N^4 k_F^{4(N-1)}} \ ,
\end{align}
where for even $N$ we should use $\beta_N =
        \sin^2( N \arccos(q/2k_F) )$, and for odd $N$ we should use $\beta_N =
        \cos^2( N \arccos(q/2k_F) )$. Interestingly, the correction to the conductivity decreases with increasing $N$. The magnetic noise for general $N > 1$ is given by
\begin{align}
    \chi_\mathrm{total}^N = 2\chi^G \left( 1 + \frac{1}{N^2} \frac{\Delta^2 + k_F^{2N} \alpha_N^2 \gamma_N}{m_0^2 \alpha_N^4 k_F^{4(N-1)}} \right) ,
\end{align}
where $\gamma_N=1$ or $\gamma_N=1/2$ for $N$ even or odd, respectively. Notice that again the spin contributions introduce the explicit appearance of the gap into the noise, in contrast to the purely orbital contributions, which are identical to the non-relativistic Galilean fermions (see Eq.\eqref{eq:N_layer_magnetic_noise}).

%\begin{align}
%    \beta_N = \begin{cases}
%        \sin^2\left( N \arccos\left( \frac{q}{2k_F} \right) \right) \ &, \ N \ \mathrm{even} \\
%        \cos^2\left( N \arccos\left( \frac{q}{2k_F} \right) \right) \ &, \ N \ \mathrm{odd} \ ,
%    \end{cases}
%\end{align}

% Notice that while we now get correction terms depending on the ratio of orbital and electron mass for the magnetic noise, our previous result of the (low-energy) Dirac fermion having twice the magnetic noise compared to the rest can still be observed (corresponding to a $2$ instead of a $1$ in the bracket).

\medskip

\textit{\bfseries \color{blue} Discussion.} 
We have analysed the transverse quasi-static conductivity of two-dimensional metals as a function of the wave-vector $q$ for a variety of systems, including parabolic and Dirac fermions as well as the chiral fermions relevant for $N$-layer rhombohedral graphene. We demonstrated that the long-wavelength limit  ($q \ll k_F$) of such transverse conductivity is universal and only controlled by the Fermi radius regardless of the presence of non-trivial Berry phases in these systems (see Figs.\ref{fig:Dirac_conductivity} and \ref{fig:multilayer_conductivity}). However, the transverse conductivity in these systems displays drastically different behaviours at finite wave-vectors. For example, when $N$ is odd (i.e. mono-layer and rhombohedrally stacked tri- and penta-layer graphene), the transverse conductivity has a divergence near the backscattering threshold ($q \to 2k_F$), while it vanishes for even $N$ as well as parabolic fermions. We also found that the conductivity has a number of zeroes equals to $N/2$ for even $N$ and $(N-1)/2$ for odd $N$. 

Interestingly, despite this non-trivial wave-vector dependence, the wave-vector integrated transverse conductivity takes a value which is independent of $N$ for $N \geq 2$, which is identical to that of Galilean fermions, but has twice this value for $N=1$ (i.e. monolayer graphene). This implies that the contribution from orbital currents to the $T_1$ time of a spin qubit placed at a distance smaller than the Fermi wavelength from the 2D sample has a value in monolayer graphene which is $1/2$ of the value of multilayer graphene or of a trivial Galilean fermion (see Eq.\eqref{eq:N_layer_magnetic_noise}). This prediction could be tested by measuring the $T_1$ times of nuclei, NV centers, or other spin qubits placed very close to the sample.

% \begin{table}[t]
%     \centering
%     \begin{tabular}{c|c  c  c  c}
%          & Galilean $\vert$ & L.E. Dirac $\vert$ & Dirac $\vert$ & Multilayer \\
%         \hline
%         Conductivity & $\sigma_\perp^G$ & $\sigma_\perp^G + M$ & $\sigma_\perp^G + M$ & $\sigma_\perp^G \cdot P_N$ \\
%         Magnetic Noise & S & 2S & 2S & S \\
%     \end{tabular}
%     \caption{Caption}
%     \label{tab:my_label}
% \end{table}

\medskip
\textit{\bfseries \color{blue} Acknowledgements.}
We are thankful to Nabeel Aslam and Jürgen Haase for stimulating discussions. This research is supported by the Deutsche Forschungsgemeinschaft (DFG) under project numbers 542614019; 518372354.

%\bibliography{transverse-conductivity}
\bibliography{PRL_draft.bib}

\clearpage

\appendix

\renewcommand{\theequation}{\thesection-\arabic{equation}}
\renewcommand{\thefigure}{\thesection-\arabic{figure}}
\renewcommand{\thetable}{\thesection-\Roman{table}}

\onecolumngrid

\section*{Supplemental Material}

\section{Derivation of the Transverse Conductivity for Galilean and Dirac Fermions}
\label{appendixA}
\noindent
Consider the general Hamiltonian in Eq.\eqref{eq:general_hamiltonian} with a transverse electric field pointing in $y$-direction. The current will be given by:
\begin{align}
    \mathbf{I} = - \nabla_\mathbf{A} H \equiv \mathbf{v}\bigr(\mathbf{p} - \mathbf{A}(\mathbf{r}, t) \bigr) \ ,
\end{align}
with the velocity operator $\mathbf{v}(\mathbf{p}) = \nabla_\mathbf{p} h(\mathbf{p})$, allowing us to write the Hamiltonian as:
\begin{align}
    H = h\bigl( \mathbf{p} - \mathbf{A}(\mathbf{r}, t) \bigr) = h(\mathbf{p}) - \int \mathbf{j}(\mathbf{r}^\prime) \cdot \mathbf{A}(\mathbf{r}^\prime, t) \, \mathrm{d}^2 r^\prime + \mathcal{O}(\mathbf{A}^2) \ ,
\end{align}
where the current density up to leading order in $A$ is given by:%\footnote{Considering higher orders in the magnetic vector potential will add a constant to the conductivity, but one finds that this constant is purely imaginary and will thus not change the real part of the transverse conductivity.}
\begin{align}
    \mathbf{j}(\mathbf{r}^\prime) = \frac{1}{2} \left\{ \mathbf{v}(\mathbf{p}) , \rho(\mathbf{r}^\prime) \right\} + \mathcal{O}(\mathbf{A}) \ .
\end{align}
Here, $\rho(\mathbf{r}^\prime) = \delta(\mathbf{r}^\prime - \mathbf{r})$ is the (single-particle) density for a particle at position $\mathbf{r}$. The linear in $\mathbf{A}$ term above will give rise to the diamagnetic current contribution, which will not affect the real part of the transverse conductivity and will therefore be ignored. The expression for the current density follows from
\begin{align}
    \mathbf{I} = \mathbf{v}\bigr(\mathbf{p} - \mathbf{A}(\mathbf{r}, t) \bigr) \overset{!}{=} \int \mathbf{j}(\mathbf{r}^\prime) \, \mathrm{d}^2 r^\prime = \frac{1}{2} \int \left\{ \mathbf{v}\bigl(\mathbf{p} - \mathbf{A}(\mathbf{r}, t) \bigr) , \rho(\mathbf{r}^\prime) \right\} \, \mathrm{d}^2 r^\prime
\end{align}
and the current density operator needing to be hermitian. Computing the Fourier transform
\begin{align}
    \mathbf{j}(\mathbf{k}) = \frac{1}{2} \left\{ \mathbf{v}(\mathbf{p}) , \mathrm{e}^{-\mathrm{i} \mathbf{k} \cdot \mathbf{r}} \right\} \ ,
\end{align}
we want to compute its matrix elements in the eigenbasis of the unperturbed Hamiltonian
\begin{align}
    H_0 = h(\mathbf{p}) \ ,
\end{align}
which is given by $\ket{\mathbf{k}} \otimes \ket{\mathbf{k}, s}$, where $\ket{\mathbf{k}}$ are the momentum eigenstates and $\ket{\mathbf{k}, s}$ denote the vectors of the matrix $h(\mathbf{k})$ for $s = \pm$. Hence, the matrix elements of the current density operator in this basis are given by
\begin{align}
    \bigr( \bra{\mathbf{k}_1} \otimes \bra{\mathbf{k}_1, s_1} \bigl) \, \mathbf{j}(\mathbf{k}) \, \bigl( \ket{\mathbf{k}_2} \otimes \ket{\mathbf{k}_2, s_2} \bigr) = \frac{1}{2} \bra{\mathbf{k}_1, s_1} \bigg( \mathbf{v}(\mathbf{k}_1) + \mathbf{v}(\mathbf{k}_2) \bigg) \ket{\mathbf{k}_2, s_2} \delta_{\mathbf{k}_1, \mathbf{k}_2 - \mathbf{k}}
\end{align}

% is a matrix, we can choose our basis as the tensor product of the momentum eigenstates and the eigenstates of the matrix, which depend on our choice of momentum. In other words, if $\mathbf{p}$ denotes the momentum operator, then we have
% \begin{align}
%     h(\mathbf{p}) \ket{\mathbf{k}} = h(\mathbf{k}) \ket{\mathbf{k}} \ ,
% \end{align}
% where $h(\mathbf{k})$ is a matrix (containing numbers) and the eigenstates of $h(\mathbf{k})$ depend on the specific choice of $\mathbf{k}$. This means we compute
% \begin{align}
%     \bra{\mathbf{k}_1, s_1(\mathbf{k}_1)} \mathbf{j}(\mathbf{k}) \ket{\mathbf{k}_2, s_2(\mathbf{k}_2)} &= \frac{1}{2} \bra{\mathbf{k}_1, s_1(\mathbf{k}_1)} \bigg( \mathbf{
%     v}(\mathbf{p}) \mathrm{e}^{-\mathrm{i} \mathbf{k} \cdot \mathbf{r}} + \mathrm{e}^{-\mathrm{i} \mathbf{k} \cdot \mathbf{r}} \mathbf{v}(\mathbf{p}) \bigg) \ket{\mathbf{k}_2, s_2(\mathbf{k}_2)} \notag \\
%     &= \frac{1}{2} \bra{s_1(\mathbf{k}_1)} \bigg( \mathbf{v}(\mathbf{k}_1) + \mathbf{v}(\mathbf{k}_2) \bigg) \ket{s_2(\mathbf{k}_2)} \bra{\mathbf{k}_1} \mathrm{e}^{-\mathrm{i} \mathbf{k} \cdot \mathbf{r}} \ket{\mathbf{k}_2} \notag \\
%     &= \frac{1}{2} \bra{s_1(\mathbf{k}_1)} \bigg( \mathbf{v}(\mathbf{k}_1) + \mathbf{v}(\mathbf{k}_2) \bigg) \ket{s_2(\mathbf{k}_2)} \delta_{\mathbf{k}_1, \mathbf{k}_2 - \mathbf{k}}
% \end{align}
% if we assume our system to have periodic boundary conditions.

Using linear response theory for a periodic potential \cite{Giuliani_Vignale_2005}, the response function w.r.t the magnetic vector potential will be given by
\begin{align}
    \chi_{yy}(q, \omega) = \frac{1}{4V} \sum_\mathbf{k} \sum_{s_1, s_2} \left\vert \bra{\mathbf{k} - \mathbf{q}, s_1} \bigg( v_y(\mathbf{k} - \mathbf{q}) + v_y(\mathbf{k}) \bigg) \ket{\mathbf{k}, s_2} \right\vert^2 \frac{n_{\mathbf{k} - \mathbf{q}, s_1} - n_{\mathbf{k}, s_2}}{\omega + \mathrm{i} \eta + \varepsilon_{\mathbf{k} - \mathbf{q}, s_1} - \varepsilon_{\mathbf{k}, s_2}}
\end{align}
% \begin{align}
%     \chi_{yy}(q, \omega) = \frac{1}{V} \sum_{\mathbf{k}_1, \mathbf{k}_2} \sum_{s_1, s_2} \bra{\mathbf{k}_1, s_1(\mathbf{k}_1)} \mathbf{j}_y(\mathbf{q}) \ket{\mathbf{k}_2, s_2(\mathbf{k}_2)} \bra{\mathbf{k}_2, s_2(\mathbf{k}_2)} \mathbf{j}_y(-\mathbf{q}) \ket{\mathbf{k}_1, s_1(\mathbf{k}_1)} \frac{n_{\mathbf{k}_1, s_1} - n_{\mathbf{k}_2, s_2}}{\omega + \mathrm{i} \eta + \varepsilon_{\mathbf{k}_1, s_1} - \varepsilon_{\mathbf{k}_2, s_2}}
% \end{align}
in the limit of $\eta \to 0$. Due our gauge choice we have
\begin{align}
    \sigma_\perp(q, \omega) = \frac{\chi_{yy}(q, \omega)}{\mathrm{i} \omega} \ ,
\end{align}
so taking the thermodynamic limit, we recover Eq.\eqref{eq:general_transverse_conductivity}.

For Galilean fermions, the Hamiltonian is not a matrix, so we do not have the sum over the index $s$. For these fermions, we also have
\begin{align}
    \mathbf{v}(\mathbf{k}) = \nabla_\mathbf{k} \varepsilon_\mathbf{k} \ .
\end{align}
Using
\begin{align}
    \mathrm{Im}\, \lim_{\eta \to 0} \frac{1}{x + \mathrm{i} \eta} = - \mathrm{i} \pi \delta(x) \ , \quad \frac{\mathrm{d} \Theta(x)}{\mathrm{d} x} = \delta(x) \ ,
\end{align}
the limits of $\eta \to 0$ and $\omega \to 0$ will give rise to two Delta functions:
\begin{align}
    \sigma_\perp^G(q) \equiv \lim_{\omega \to 0} \mathrm{Re}\, \sigma_\perp^G(q, \omega) = \frac{\pi}{4} \int \frac{\mathrm{d}^2 k}{(2\pi)^2} \bigg( v_y(\mathbf{k} - \mathbf{q}) + v_y(\mathbf{k}) \bigg)^2 \delta(\varepsilon_F - \varepsilon_\mathbf{k}) \delta(\varepsilon_{\mathbf{k} - \mathbf{q}} - \varepsilon_\mathbf{k}) \ .
\end{align}
With a suitable coordinate transform, one can find rewrite this as
\begin{align}\label{eq:appendix:Galilean_conductivity_raw}
    \sigma_\perp^G(q) = \frac{1}{16\pi} \sum_i \frac{\bigl( v_y(\mathbf{k}_i - \mathbf{q}/2) + v_y(\mathbf{k}_i + \mathbf{q}/2) \bigr)^2}{\vert \det J(\mathbf{k}_i) \vert} \ ,
\end{align}
where we sum over ``hotspots''
\begin{align}
    \{ \mathbf{k}_i \} = \{ \mathbf{k} \in \mathbb{R}^2 \ \vert \ \varepsilon_F = \varepsilon_{\mathbf{k}_i - \mathbf{q}/2} \ , \ \varepsilon_{\mathbf{k}_i - \mathbf{q}/2} = \varepsilon_{\mathbf{k}_i + \mathbf{q}/2} \} \ .
\end{align}
For an isotropic dispersion, we find
\begin{align}\label{eq:appendix:isotropic_hotspots}
    \{ \mathbf{k}_i \} = \left\{ \pm \hat{\mathbf{y}} \sqrt{k_F^2 - \frac{q^2}{4}} \right\} \ ,
\end{align}
allowing us to write
\begin{align}
    v_x(\mathbf{k}_i - \mathbf{q}/2) = -v_x(\mathbf{k}_i + \mathbf{q}/2) \ , \quad v_y(\mathbf{k}_i - \mathbf{q}/2) = v_y(\mathbf{k}_i + \mathbf{q}/2) \ ,
\end{align}
resulting in the Jacobian determinant
\begin{align}
    \vert \det J(\mathbf{k}_i) \vert = 2 v_x(\mathbf{k}_i + \mathbf{q}/2) v_y(\mathbf{k}_i + \mathbf{q}/2) \ .
\end{align}
The conductivity of the Galilean fermion therefore becomes
\begin{align}
    \mathrm{Re}\, \sigma_\perp^G(q) = \frac{1}{8\pi} \sum_i \left\vert \frac{v_y(\mathbf{k}_i + \mathbf{q}/2)}{v_x(\mathbf{k}_i + \mathbf{q}/2)} \right\vert = \begin{cases}
        \frac{1}{2\pi} \frac{k_F}{q} \sqrt{1 - \frac{q^2}{4k_F^2}} \ &, \ q\leq 2k_F \\
        0 \ &, \ q > 2k_F \ .
    \end{cases}
\end{align}

For the Dirac fermion, we find
\begin{align}
    \mathbf{v}(\mathbf{k}) = v \boldsymbol{\tau} \ , \quad \varepsilon_{\mathbf{k},s} = s \varepsilon_\mathbf{k} = s \sqrt{v^2 \vert \mathbf{k} \vert^2 + \Delta^2} \ , \quad n_{\mathbf{k}, s} = \begin{cases}
        n_\mathbf{k} \ &, \ s = + \\
        1 \ &, \ s = -
    \end{cases} \ , \quad n_\mathbf{k} = \Theta(\varepsilon_F - \varepsilon_\mathbf{k}) \ .
\end{align}
Inserting this into Eq.\eqref{eq:general_transverse_conductivity} and taking the limits $\eta \to 0$ and $\omega \to 0$, we find that only $s_1 = s_2 = +$ will contribute to the conductivity. By similar argument to the Galilean fermion we therefore get
\begin{align}
    \sigma_\perp^D(q) \equiv \lim_{\omega \to 0} \mathrm{Re} \, \sigma_\perp^D(q, \omega) = \frac{v^2}{4\pi} \sum_i \frac{\left\vert \bra{\mathbf{k}_i - \mathbf{q}/2, +} \tau_y \ket{\mathbf{k}_i + \mathbf{q}/2, +} \right\vert^2}{\left\vert \det J(\mathbf{k}_i) \right\vert} \ ,
\end{align}
with identical hotspots as for the Galilean fermion given by Eq.\eqref{eq:appendix:isotropic_hotspots} and Jacobian determinant
\begin{align}
    \vert \det J(\mathbf{k}_i) \vert = \frac{v^4 qk_F}{\varepsilon_F^2} \sqrt{1 - \frac{q^2}{4k_F^2}} \ .
\end{align}
Using
\begin{align}\label{eq:appendix:trace_trick}
    \vert \bra{\mathbf{k}_1, +} \tau_j \ket{\mathbf{k}_2, +} \vert^2 &= \mathrm{tr} \bigg( \tau_j \ket{\mathbf{k}_1, +}\bra{\mathbf{k}_1, +} \tau_j \ket{\mathbf{k}_2, +} \bra{\mathbf{k}_2, +} \bigg) \notag \\
    &= \mathrm{tr} \left( \tau_j \frac{\mathbb{1} + \sum_m \hat{d}_m(\mathbf{k}_1) \tau_m}{2} \tau_j \frac{\mathbb{1} + \sum_n \hat{d}_n(\mathbf{k}_2) \tau_n}{2} \right) \ ,
\end{align}
where we define
\begin{align}
    h(\mathbf{k}) = \sum_m d_m(\mathbf{k}) \tau_m \ , \quad \hat{\mathbf{d}}(\mathbf{k}) = \frac{\mathbf{d}(\mathbf{k})}{\vert \mathbf{d}(\mathbf{k}) \vert} \equiv \frac{1}{\varepsilon_\mathbf{k}} \begin{pmatrix}
        v k_x \\ v k_y \\ \Delta
    \end{pmatrix} \ ,
\end{align}
gives us
\begin{align}
    \vert \bra{\mathbf{k}_i - \mathbf{q}/2, +} \tau_y \ket{\mathbf{k}_i + \mathbf{q}/2, +} \vert^2 = \frac{v^2 k_F^{2}}{\varepsilon_F^2} \ .
\end{align}
This results in
\begin{align}
    \sigma_\perp^D(q) = \frac{\Theta(2kF - q)}{2\pi} \frac{k_F}{q} \frac{1}{\sqrt{1 - \frac{q^2}{4k_F^2}}} = \frac{\Theta(2k_F - q)}{2\pi} \frac{k_F}{q} \sqrt{1 - \frac{q^2}{4k_F^2}} + \frac{\Theta(2k_F - q)}{8\pi} \frac{q}{k_F} \frac{1}{\sqrt{1 - \frac{q^2}{4k_F^2}}} \ .
\end{align}

\section{Correction to the Conductivity due to Orbital Magnetic Moment}
\label{appendixB}
\noindent
Take a Hamiltonian with a magnetic moment $\mu$ coupling to the $z$-component of the magnetic field $B_z(\mathbf{r}, t)$, similar to the Hamiltonian in Eq.$\eqref{eq:orbital_moment_hamiltonian}$:
\begin{align}
    H = \frac{\bigl( \mathbf{p} - \mathbf{A}(\mathbf{r}, t) \bigr)^2}{2m} + \mu B_z(\mathbf{r}, t) \ .
\end{align}
We can use integration by parts to write
\begin{align}
    \mu B_z(\mathbf{r}, t) = \mu \bigl( \nabla \times \mathbf{A}(\mathbf{r}, t) \bigr)_z = -\mu \int \bigg( A_y(\mathbf{r}^\prime, t) \partial_{x^\prime} \rho(\mathbf{r}^\prime) - A_x(\mathbf{r}^\prime, t) \partial_{y^\prime} \rho(\mathbf{r}^\prime) \bigg) \, \mathrm{d}^2 \mathbf{r}^\prime \ ,
\end{align}
where $\rho(\mathbf{r}^\prime) = \delta(\mathbf{r}^\prime - \mathbf{r})$ is again the (single-particle) density of a particle at position $\mathbf{r}$. Following similar steps as in Appendix \ref{appendixA}, this term gives rise to a current density operator
\begin{align}\label{eq:appendix:magnetic_moment_current}
    \mathbf{j}_\mu(\mathbf{r}^\prime) = -\mu \bigg( \partial_{y^\prime} \rho(\mathbf{r}^\prime) \mathbf{\hat{x}} - \partial_{x^\prime} \rho(\mathbf{r}^\prime) \mathbf{\hat{y}} \bigg) \ ,
\end{align}
allowing us to write the Hamiltonian up to leading order in the magnetic vector potential as
\begin{align}
    H = h(\mathbf{p}) - \int \bigg( \mathbf{j}(\mathbf{r}^\prime) + \mathbf{j}_\mu(\mathbf{r}^\prime) \bigg) \cdot \mathbf{A}(\mathbf{r}^\prime, t) \, \mathrm{d}^2 r^\prime + \mathcal{O}(\mathbf{A}^2) \ ,
\end{align}
where $\mathbf{j}$ is the current density of the Galilean fermion. The total transverse conductivity is then given by
\begin{align}\label{eq:appendix:LED:mix_terms}
    \sigma_\perp^{LD}(q, \omega) = \frac{1}{\mathrm{i} \omega} \bigg( \chi_{j_y j_y}(q, \omega) + \chi_{j_y j_{\mu, y}}(q, \omega) + \chi_{j_{\mu, y} j_y}(q, \omega) + \chi_{j_{\mu, y} j_{\mu, y}}(q, \omega) \bigg) \ ,
\end{align}
with general response function
% \begin{align}
%     \chi_{AB} = \frac{1}{V} \sum_{\mathbf{k}} \sum_{s_1, s_2} A_{\mathbf{k} - \mathbf{q}, s_1, \mathbf{k}, s_2}(\mathbf{q}) B_{\mathbf{k}, s_2, \mathbf{k}-\mathbf{q}, s_1}(-\mathbf{q}) \frac{n_{\mathbf{k}- \mathbf{q}, s_1} - n_{\mathbf{k}, s_2}}{\omega + \mathrm{i} \eta + \varepsilon_{\mathbf{k} - \mathbf{q}, s_1} - \varepsilon_{\mathbf{k}, s_2}} \ ,
% \end{align}
% where we set
% \begin{align}
%     \varepsilon_{\mathbf{k}, s} = s \varepsilon_\mathbf{k} \ , \quad \varepsilon_\mathbf{k} = \sqrt{v^2 \vert \mathbf{k} \vert^2 + \Delta^2} \ , \quad n_{\mathbf{k}, s} = \begin{cases}
%         n_\mathbf{k} \ &, \ s = + \\
%         1 \ &, \ s = -
%     \end{cases} \ , \quad n_\mathbf{k} = \Theta(\varepsilon_F - \varepsilon_\mathbf{k})
% \end{align}
% and
% \begin{align}
%     A_{\mathbf{k}_1, s_1, \mathbf{k}_2, s_2}(\mathbf{q}) \equiv \bigl( \bra{\mathbf{k}_1} \otimes \bra{\mathbf{k}_1, s_1} \bigr) \, A(\mathbf{q}) \, \bigl( \ket{\mathbf{k}_2} \otimes \ket{\mathbf{k}_2, s_2} \bigr) \ .
% \end{align}
\begin{align}\label{eq:appendix:general_linear_response_function}
    \chi_{AB} = \int \frac{\mathrm{d}^2 k}{(2\pi)^2} \bra{\mathbf{k}- \mathbf{q}} A(\mathbf{q}) \ket{\mathbf{k}} \bra{\mathbf{k}} B(-\mathbf{q}) \ket{\mathbf{k} - \mathbf{q}} \frac{n_{\mathbf{k}- \mathbf{q}} - n_{\mathbf{k}}}{\omega + \mathrm{i} \eta + \varepsilon_{\mathbf{k} - \mathbf{q}} - \varepsilon_{\mathbf{k}}} \ ,
\end{align}
where we set
\begin{align}
    \varepsilon_\mathbf{k} = \frac{\mathbf{k}^2}{2m} \ , \quad n_\mathbf{k} = \Theta(\varepsilon_F - \varepsilon_\mathbf{k}) \ .
\end{align}
The first term in Eq.\eqref{eq:appendix:LED:mix_terms}, $\chi_{j_y j_y} \equiv \chi_{yy}$, corresponds to the Galilean conductivity calculated in Appendix \ref{appendixA}. The cross-current terms cancel each other, i.e.
\begin{align}
    \chi_{j_y j_{\mu, y}}(q, \omega) + \chi_{j_{\mu, y} j_y}(q, \omega) = 0 \ ,
\end{align}
so it remains to compute the final term, which by similar calculation to Appendix \ref{appendixA} can be found to be
\begin{align}\label{eq:appendix:mu_mu_current_correction}
    \lim_{\omega \to 0} \mathrm{Re}\, \frac{\chi_{j_{\mu, y} j_{\mu, y}}(q, \omega)}{\mathrm{i} \omega} = \frac{\mu^2 q^2}{8\pi} \sum_i \frac{1}{\vert v_x(\mathbf{k}_i + \mathbf{q}/2) v_y(\mathbf{k}_i + \mathbf{q}/2) \vert} \ ,
\end{align}
where $v_j$ is the $j$-th component of the Galilean velocity operator, and we sum over the same hotspots as before, given in Eq.\eqref{eq:appendix:isotropic_hotspots}. Inserting
\begin{align}
    \mu = -\frac{1}{2m} \ , \quad v_j(\mathbf{k}) = \frac{k_j}{m} \ ,
\end{align}
we arrive at
\begin{align}
    \sigma_\perp^{LD}(q) = \frac{\Theta(2k_F - q)}{2\pi} \frac{k_F}{q} \sqrt{1 - \frac{q^2}{4k_F^2}} + \frac{\Theta(2k_F - q)}{8\pi} \frac{q}{k_F} \frac{1}{\sqrt{1 - \frac{q^2}{4k_F^2}}} \ ,
\end{align}
which is exaclty identical to the Dirac conductivity in Eq.\eqref{eq:Dirac_conductivity}.

\section{Derivation of the Multilayer Conductivity}
\label{appendixC}
\noindent
Using the spin lowering and raising operators
\begin{align}
    \tau_\pm = \tau_x \pm \mathrm{i} \tau_y \quad \implies \quad \begin{cases}
        \tau_x = \frac{\tau_+ + \tau_-}{2} \\
        \tau_y = \frac{\tau_+ - \tau_-}{2\mathrm{i}}
    \end{cases}
\end{align}
we can rewrite our Hamiltonian in Eq.\eqref{eq:multilayer_hamiltonian} as
\begin{align}
    h(\mathbf{p}) &= \alpha_N \begin{pmatrix} \mathrm{Re}\, (p_x + \mathrm{i} p_y)^N \\ \mathrm{Im}\, (p_x + \mathrm{i} p_y)^N \end{pmatrix} \cdot \boldsymbol{\tau} + \Delta \tau_z = \frac{\alpha_N}{2} \bigg( (p_x - \mathrm{i} p_y)^N \tau_+ + (p_x + \mathrm{i} p_y)^N \tau_- \bigg) + \Delta \tau_z \ ,
\end{align}
where we set $\boldsymbol{\tau} = (\tau_x, \tau_y)$. The $y$-component of the velocity operator is hence given by:
\begin{align}
    v_y(\mathbf{k}) = \frac{\partial h(\mathbf{k})}{\partial k_y} = \frac{N \alpha_N}{2} \bigg( -\mathrm{i} (k_x - \mathrm{i} k_y)^{N-1} \tau_+ + \mathrm{i} (k_x + \mathrm{i} k_y)^{N-1} \tau_- \bigg) = N \alpha_N \begin{pmatrix} - \mathrm{Im}\, (k_x + \mathrm{i} k_y)^{N-1} \\  \mathrm{Re}\, (k_x + \mathrm{i} k_y)^{N-1} \end{pmatrix} \cdot \boldsymbol{\tau} \ ,
\end{align}
% Similar as in Appendix \ref{appendixA}, we can write
% \begin{align}
%     H = h(\mathbf{p}) - \int \mathbf{j}(\mathbf{r}^\prime) \cdot \mathbf{A}(\mathbf{r}^\prime, t) \diff^2 r^\prime + \mathcal{O}(\mathbf{A}^2) \ , \quad \mathbf{j}(\mathbf{r}^\prime) = \frac{1}{2} \left\{ \mathbf{v}(\mathbf{p}) , \rho(\mathbf{r}^\prime) \right\} \ ,
% \end{align}
% so following similar steps gives us
which we can insert into Eq.\eqref{eq:general_transverse_conductivity}, where we have
\begin{align}
    \varepsilon_{\mathbf{k}, s} = s \varepsilon_\mathbf{k} \ , \quad n_{\mathbf{k}, s} = \begin{cases}
        n_\mathbf{k} \ &, \ s = + \\
        1 \ &, \ s = -
    \end{cases}
\end{align}
with
\begin{align}
    \varepsilon_\mathbf{k} = \sqrt{\alpha_N^2 \mathbf{k}^{2N} + \Delta^2} \ , \quad n_\mathbf{k} = \Theta(\varepsilon_F - \varepsilon_\mathbf{k}) \ .
\end{align}
%Due to
% \begin{align}
%     \mathrm{Im}\, \lim_{\eta \to 0} \frac{1}{x + \mathrm{i} \eta} = - \mathrm{i} \pi \delta(x) \ , \quad \frac{\mathrm{d} \Theta(x)}{\mathrm{d} x} = \delta(x) \ ,
% \end{align}
% taking the limit of $\eta \to 0$ and $\omega \to 0$ will give rise to two Delta functions:
% \begin{align}
%     \sigma_\perp^N(q) \equiv \lim_{\omega \to 0} \mathrm{Re}\, \sigma_\perp^N(q, \omega) = -\frac{\pi}{4}
% \end{align}
% Using
% \begin{align}
%     \mathrm{Im}\, \lim_{\eta \to 0} \frac{1}{x + \mathrm{i} \eta} = - \mathrm{i} \pi \delta(x) \ , \quad \frac{\mathrm{d} \Theta(x)}{\mathrm{d} x} = \delta(x) \ ,
% \end{align}
% the limits of $\eta \to 0$ and $\omega \to 0$ will give rise to two Delta functions, which cause that only the case of $s_1 = s_2 = +$ will contribute to the conductivity:
% \begin{align}
%     \sigma_\perp^N(q) \equiv \lim_{\omega \to 0} \mathrm{Re}\, \sigma_\perp^N(q, \omega) = \frac{\pi}{4} \int \frac{\mathrm{d}^2 k}{(2\pi)^2} \left\vert \bra{\mathbf{k} - \mathbf{q}, +} \bigg( v_y(\mathbf{k} - \mathbf{q}) + v_y(\mathbf{k}) \bigg) \ket{\mathbf{k}, +} \right\vert^2 \delta(\varepsilon_F - \varepsilon_\mathbf{k}) \delta(\varepsilon_{\mathbf{k} - \mathbf{q}} - \varepsilon_\mathbf{k}) \ .
% \end{align}
% With a suitable coordinate transform, one can find rewrite this as
By similar calculation to Appendix \ref{appendixA}, we arrive at
\begin{align}\label{eq:appendix:multilayer_conductivity_raw}
    \sigma_\perp^N(q) \equiv \lim_{\omega \to 0} \mathrm{Re}\, \sigma_\perp^N(q, \omega) = \frac{1}{16\pi} \sum_i \frac{\bra{\mathbf{k}_i - \mathbf{q}/2, +} \bigl( v_y(\mathbf{k}_i - \mathbf{q}/2) + v_y(\mathbf{k}_i + \mathbf{q}/2) \bigr) \ket{\mathbf{k}_i + \mathbf{q}_2, +}}{\vert \det J(\mathbf{k}_i) \vert} \ ,
\end{align}
where we sum over the same hotspots given by Eq.\eqref{eq:appendix:isotropic_hotspots}
% where we sum over ``hotspots''
% \begin{align}
%     \{ \mathbf{k}_i \} = \{ \mathbf{k} \in \mathbb{R}^2 \ \vert \ \varepsilon_F = \varepsilon_{\mathbf{k}_i - \mathbf{q}/2} \ , \ \varepsilon_{\mathbf{k}_i - \mathbf{q}/2} = \varepsilon_{\mathbf{k}_i + \mathbf{q}/2} \} \ .
% \end{align}
% For an isotropic dispersion, we find
% \begin{align}
%     \{ \mathbf{k}_i \} = \left\{ \pm \hat{\mathbf{y}} \sqrt{k_F^2 - \frac{q^2}{4}} \right\}
% \end{align}
with Jacobian determinant
\begin{align}\label{eq:appendix:multilayer_Jacobian}
    \vert \det J(\mathbf{k}_i) \vert = \frac{N^2 \alpha_N^4 k_F^{4N} q}{k_F^3 \varepsilon_F^2} \sqrt{1 - \frac{q^2}{4k_F^2}} \ .
\end{align}
It remains to compute the matrix elements of the velocity operator. Since the hotspots lie along the $y$-axis, for odd $N \in \mathbb{N}$ we find
\begin{align}
    \mathrm{Re}\, \left( - \frac{q}{2} + \mathrm{i} k_{i, y} \right)^N = -\mathrm{Re}\, \left( + \frac{q}{2} + \mathrm{i} k_{i, y} \right)^N \ , \quad \mathrm{Im}\, \left( - \frac{q}{2} + \mathrm{i} k_{i, y} \right)^N = \mathrm{Im}\, \left( + \frac{q}{2} + \mathrm{i} k_{i, y} \right)^N
\end{align}
and similarly for even $N \in \mathbb{N}$:
\begin{align}
    \mathrm{Re}\, \left( - \frac{q}{2} + \mathrm{i} k_{i, y} \right)^N = \mathrm{Re}\, \left( + \frac{q}{2} + \mathrm{i} k_{i, y} \right)^N \ , \quad \mathrm{Im}\, \left( - \frac{q}{2} + \mathrm{i} k_{i, y} \right)^N = -\mathrm{Im}\, \left( + \frac{q}{2} + \mathrm{i} k_{i, y} \right)^N \ ,
\end{align}
allowing us to write
\begin{align}\label{eq:appendix:multilayer_velocity_matrix_elements}
    \vert \bra{\mathbf{k}_i - \mathbf{q}/2} \Bar{v}_y \ket{\mathbf{k}_i + \mathbf{q}/2} \vert^2 = \begin{cases}
        N^2 \alpha_N^2 \left( \mathrm{Re}\, \left( \frac{q}{2} + \mathrm{i} k_{i, y} \right)^{N-1} \right)^2 \vert \bra{\mathbf{k}_i - \mathbf{q}/2} \tau_y \ket{\mathbf{k}_i + \mathbf{q}/2} \vert^2 \ &, \ N \ \mathrm{odd} \\
        N^2 \alpha_N^2 \left( \mathrm{Im}\, \left( \frac{q}{2} + \mathrm{i} k_{i, y} \right)^{N-1} \right)^2 \vert \bra{\mathbf{k}_i - \mathbf{q}/2} \tau_x \ket{\mathbf{k}_i + \mathbf{q}/2} \vert^2 \ &, \ N \ \mathrm{even} \\
    \end{cases}
\end{align}
where we defined the short-hand
\begin{align}
    \Bar{v}_y := \frac{1}{2} \bigg( v_y(\mathbf{k}_i - \mathbf{q}/2) + v_y(\mathbf{k}_i + \mathbf{q}/2) \bigg) \ .
\end{align}
Using Eq.\eqref{eq:appendix:trace_trick}
% \begin{align}
%     \vert \bra{\mathbf{k}_1, +} \tau_j \ket{\mathbf{k}_2, +} \vert^2 &= \mathrm{tr} \bigg( \tau_j \ket{\mathbf{k}_1, +}\bra{\mathbf{k}_1, +} \tau_j \ket{\mathbf{k}_2, +} \bra{\mathbf{k}_2, +} \bigg) \notag \\
%     &= \mathrm{tr} \left( \tau_j \frac{\mathbb{1} + \sum_m \hat{d}_m(\mathbf{k}_1) \tau_m}{2} \tau_j \frac{\mathbb{1} + \sum_n \hat{d}_n(\mathbf{k}_2) \tau_n}{2} \right) \ ,
% \end{align}
% where we define
with
\begin{align}
    h(\mathbf{k}) = \sum_m d_m(\mathbf{k}) \tau_m \ , \quad \hat{\mathbf{d}}(\mathbf{k}) = \frac{\mathbf{d}(\mathbf{k})}{\vert \mathbf{d}(\mathbf{k}) \vert} \equiv \frac{1}{\varepsilon_\mathbf{k}} \begin{pmatrix}
        \alpha_N \mathrm{Re}\, (k_x + \mathrm{i} k_y)^N \\ \alpha_N \mathrm{Im} \, (k_x + \mathrm{i} k_y)^N \\ \Delta
    \end{pmatrix} \ ,
\end{align}
we find
% gives us
% \begin{align}
%     \vert \bra{\mathbf{k}_1} \tau_x \ket{\mathbf{k}_2} \vert^2 &= \frac{1}{2} \bigg( 1 + \hat{d}_x(\mathbf{k}_1) \hat{d}_x(\mathbf{k}_2) - \hat{d}_y(\mathbf{k}_1) \hat{d}_y(\mathbf{k}_2) - \hat{d}_z(\mathbf{k}_1) \hat{d}_z(\mathbf{k}_2) \bigg) \\
%     \vert \bra{\mathbf{k}_1} \tau_y \ket{\mathbf{k}_2} \vert^2 &= \frac{1}{2} \bigg( 1 - \hat{d}_x(\mathbf{k}_1) \hat{d}_x(\mathbf{k}_2) + \hat{d}_y(\mathbf{k}_1) \hat{d}_y(\mathbf{k}_2) - \hat{d}_z(\mathbf{k}_1) \hat{d}_z(\mathbf{k}_2) \bigg)
% \end{align}
\begin{align}
    \vert \bra{\mathbf{k}_i - \mathbf{q}/2, +} \tau_x \ket{\mathbf{k}_i + \mathbf{q}/2, +} \vert^2 = \vert \bra{\mathbf{k}_i - \mathbf{q}/2, +} \tau_y \ket{\mathbf{k}_i + \mathbf{q}/2, +} \vert^2 = \frac{\alpha_N^2 k_F^{2N}}{\varepsilon_F^2} \ .
\end{align}
Defining
\begin{align}
    \frac{q}{2} + \mathrm{i} \, \mathrm{sgn}(k_{i, y}) \sqrt{k_F^2 + \frac{q^2}{4}} =: k_F \mathrm{e}^{\mathrm{i} \phi} \ , \quad \phi = \mathrm{sgn}(k_{i, y}) \arccos \frac{q}{2k_F} \ ,
\end{align}
we can write
\begin{align}\label{eq:appendix:polynomials}
    \left( \mathrm{Re}\, \left( \frac{q}{2} + \mathrm{i} k_{i, y} \right)^{N-1} \right)^2 = k_F^{2(N-1)} \cos^2 \bigl( (N-1) \phi \bigr) \ , \quad \left( \mathrm{Im}\, \left( \frac{q}{2} + \mathrm{i} k_{i, y} \right)^{N-1} \right)^2 = k_F^{2(N-1)} \sin^2 \bigl( (N-1) \phi \bigr) \ ,
\end{align}
which both vanish exactly if
\begin{align}\label{eq:appendix:multilayer_zeros}
    \frac{q}{k_F} = 2\sin\left( \frac{(2m + 1) \pi}{2(N-1)} \right)
\end{align}
for $m = 0, \dots,  N-2$. Since both terms in Eq.\eqref{eq:appendix:polynomials} are polynomials in $q$, we can write them by factoring their zeros multiplied by their leading coefficient. We eventually find for odd $N> 1$
\begin{align}
    \frac{\left( \mathrm{Re}\, \left( \frac{q}{2} + \mathrm{i} k_{i,y} \right)^{N-1} \right)^2}{k_F^{2(N-1)}} = \frac{1}{4} \prod_{m=0}^{N-2} \left( \frac{q^2}{k_F^2} - 4 \sin^2 \left( \frac{(2m+1) \pi}{2(N-1)} \right) \right) 
\end{align}
and for even $N$
\begin{align}
    \frac{\left( \mathrm{Im}\, \left( \frac{q}{2} + \mathrm{i} k_{i,y} \right)^{N-1} \right)^2}{k_F^{2(N-1)}} = \frac{1}{4} \prod_{m=0}^{N-2} \left( \frac{q^2}{k_F^2} - 4 \sin^2 \left( \frac{(2m+1) \pi}{2(N-1)} \right) \right) \ .
\end{align}
Inserting everything into Eq.\eqref{eq:appendix:multilayer_conductivity_raw}, we recover Eq.\eqref{eq:multilayer_conductivity}. 

The condctivitiy is zero if the matrix elements of the $y$-component of the velocity (see Eq.\eqref{eq:appendix:multilayer_velocity_matrix_elements}) vanish. The zeros, given by Eq.\eqref{eq:appendix:multilayer_zeros}, can be written in the form of Eq.\eqref{eq:multilayer_zeros}, which allows us to visualise them by writing
\begin{align}
    v_y(\mathbf{k}) = \boldsymbol{\xi}(\mathbf{k}) \cdot \boldsymbol{\tau} \ , \quad \boldsymbol{\xi}(\mathbf{k}) = N \alpha_N \begin{pmatrix}
        -\mathrm{Im}\, (k_x + \mathrm{i} k_y)^{N-1} \\ \mathrm{Re}\, (k_x + \mathrm{i} k_y)^{N-1}
    \end{pmatrix}
\end{align}
and looking the behaviour of $\boldsymbol{\xi}$ between the two scattered states (see inset of Fig. \ref{fig:multilayer_conductivity}). We find the conductivity vanishes exactly if:
\begin{align}
    \left \vert \bra{\mathbf{k}_1} \Bar{v}_y \ket{\mathbf{k}_2} \right\vert^2 = 0 \quad \iff \quad \boldsymbol{\xi}(\mathbf{k}_1) = - \boldsymbol{\xi}(\mathbf{k}_2) \ ,
\end{align}
where $\mathbf{k}_2 = \mathbf{k}_1 + \mathbf{q}$. Conversely, we find that the conductivity is ``maximal'', i.e. it is equal to the Dirac conductivity, exactly if $\boldsymbol{\xi}$ aligns for both states, i.e. if
\begin{align}
    \boldsymbol{\xi}(\mathbf{k}_1) = \boldsymbol{\xi}(\mathbf{k}_2) \ .
\end{align}

\section{Correction to the Conductivity due to Spin Magnetic Moment}
\label{appendixD}
\noindent
Introducing the electron spin will add a correction to the Hamiltonian of form Eq.\eqref{eq:vacuum_moment_hamiltonian_correction}:
\begin{align}
    H = \begin{pmatrix}
        h\bigl( \mathbf{p} - \mathbf{A}(\mathbf{r}, t) \bigr) - \mu_s B_z(\mathbf{r}, t) & 0 \\
        0 & h\bigl( \mathbf{p} - \mathbf{A}(\mathbf{r}, t) \bigr) + \mu_s B_z(\mathbf{r}, t)
    \end{pmatrix} \ ,
\end{align}
where $\mu_s = 1/(2m_0)$ is the spin magnetic moment. For the Galilean fermion, the calculation is effectively identical to the one in Appendix \ref{appendixB} when replacing $\mu = -\mu_s$, with Eq.\eqref{eq:appendix:general_linear_response_function} now including sums over spin eigenvalues due to the additional degree of freedom. In other words, the resulting conductivity is given by Eq.\eqref{eq:appendix:LED:mix_terms} (the cross-current terms still vanish):
\begin{align}\label{eq:appendix:Galilean_spin:total}
    \sigma_\perp^{G, \mathrm{total}}(q, \omega) = \frac{1}{\mathrm{i} \omega} \bigg( \chi_{j_y j_y}(q, \omega) + \chi_{j_{\mu ,y} j_{\mu, y}}(q, \omega) \bigg) \ ,
\end{align}
where we set
\begin{align}\label{eq:appendix:LRT_function}
    \chi_{AB}(q, \omega) = \int \frac{\mathrm{d}^2 k}{(2\pi)^2} \sum_{s_1, s_2} A_{\mathbf{k} - \mathbf{q}, \mathbf{k}, s_1, s_2}(\mathbf{q}) B_{\mathbf{k}, \mathbf{k} - \mathbf{q}, s_2, s_1}(-\mathbf{q}) \frac{n_{\mathbf{k} - \mathbf{q}, s_1} - n_{\mathbf{k}, s_2}}{\omega + \mathrm{i} \eta + \varepsilon_{\mathbf{k} - \mathbf{q}, s_1} - \varepsilon_{\mathbf{k}, s_2}} \ ,
\end{align}
with $s_{1,2} = \uparrow, \downarrow$ and
\begin{align}
    A_{\mathbf{k} - \mathbf{q}, \mathbf{k}, s_1, s_2} \equiv \bigl( \bra{\mathbf{k} - \mathbf{q}} \otimes \bra{s_1} \bigr) A \bigl( \ket{\mathbf{k}} \otimes \ket{s_2} \bigr) \ , \quad n_{\mathbf{k}, s} = n_\mathbf{k} \ , \quad \varepsilon_{\mathbf{k}, s} = \varepsilon_\mathbf{k} \ .
\end{align}
The first term in Eq.\eqref{eq:appendix:Galilean_spin:total} again corresponds to the Galilean conductivity, with the additional degree of freedom giving us an additional factor $2$. The second term can be computed in a similar manner as in Appendix \ref{appendixB} (see Eq.\eqref{eq:appendix:mu_mu_current_correction}) and corresponds to the term in Eq.\eqref{eq:Galilean:spin_correction}.
% Hence, the first term in Eq.\eqref{eq:appendix:LED:mix_terms} is twice the spinless Galilean conductivity, the cross-current terms vanish, and

The exact same calculation can be done for the low-energy Dirac fermion, we simply need to add the term $\mu_\mathrm{orbital} B_z(\mathbf{r}, t) \mathbb{1}$ to the Galilean Hamiltonian, where $\mu_\mathrm{orbital}$ is the orbital magnetic moment arising from the Berry phase.

For the Dirac and chiral fermion, we have an additional degree of freedom, so in Eq.\eqref{eq:appendix:LRT_function} we need to include a sum over the pseudospin states:
\begin{align}
    \chi_{AB}(q, \omega) = \int \frac{\mathrm{d}^2 k}{(2\pi)^2} \sum_{s_1, s_2} \sum_{t_1, t_2} A_{\mathbf{k} - \mathbf{q}, \mathbf{k}, s_1, s_2, t_1, t_2}(\mathbf{q}) B_{\mathbf{k}, \mathbf{k} - \mathbf{q}, s_2, s_1, t_1, t_2}(-\mathbf{q}) \frac{n_{\mathbf{k} - \mathbf{q}, s_1, t_1} - n_{\mathbf{k}, s_2, t_2}}{\omega + \mathrm{i} \eta + \varepsilon_{\mathbf{k} - \mathbf{q}, s_1, t_2} - \varepsilon_{\mathbf{k}, s_2, t_2}} \ ,
\end{align}
with $s_{1,2} = \uparrow, \downarrow$, $t_{1,2} = \pm$, and
\begin{align}
    A_{\mathbf{k} - \mathbf{q}, \mathbf{k}, s_1, s_2, t_1, t_2} \equiv \bigl( \bra{\mathbf{k} - \mathbf{q}} \otimes \bra{s_1} \otimes \bra{\mathbf{k} - \mathbf{q}, t_1} \bigr) A \bigl( \ket{\mathbf{k}} \otimes \ket{s_2} \otimes \ket{\mathbf{k}, t_2} \bigr) \ , \quad n_{\mathbf{k}, s, t} = \begin{cases}
        n_\mathbf{k} &, \ t = + \\
        1 &, \ t = -
    \end{cases} \ , \quad \varepsilon_{\mathbf{k}, s, t} = t\varepsilon_\mathbf{k} \ .
\end{align}
Again, there will be two terms contributing to the conductivity (analogous to Eq.\eqref{eq:appendix:Galilean_spin:total}): The first one will be the conductivity computed without spin (with an additional factor $2$), while for $N \geq 1$ the other can be computed similar as in Appendix \ref{appendixA} to be:
\begin{align}
    \delta \sigma_\perp^{N, \mathrm{spin}}(q, \omega) = \frac{\mu_s^2 q^2}{2\pi} \sum_i \frac{\vert \braket{\mathbf{k}_i - \mathbf{q}/2, + \vert \mathbf{k}_i + \mathbf{q}/2, +} \vert^2}{\vert \det J(\mathbf{k}_i) \vert} \ ,
\end{align}
with hotspots and Jacobian determinant given by Eq.\eqref{eq:appendix:isotropic_hotspots} and Eq.\eqref{eq:appendix:multilayer_Jacobian} respectively. Similar to Eq.\eqref{eq:appendix:trace_trick}, we can determine the numerator to be
\begin{align}
    \left\vert \braket{\mathbf{k}_i - \mathbf{q}/2, + \vert \mathbf{k}_i + \mathbf{q}/2, +} \right\vert^2 = \begin{cases}
        1 - \frac{\alpha_N^2 \left( \mathrm{Im}\, \left( \frac{q}{2} + \mathrm{i} k_{i, y} \right) \right)^2}{\varepsilon_F} \ &, \ N \ \mathrm{even} \\
        1 - \frac{\alpha_N^2 \left( \mathrm{Re}\, \left( \frac{q}{2} + \mathrm{i} k_{i, y} \right) \right)^2}{\varepsilon_F} \ &, \ N \ \mathrm{odd} \ ,
    \end{cases}
\end{align}
leading to Eq.\eqref{eq:multilayer_spin_correction}. We get the result for the Dirac fermion by setting $N = 1$.

\end{document}